\documentclass{article}
\usepackage{spconf,graphicx}
\usepackage{multirow}
\usepackage[table,xcdraw,table,x11names, svgnames, dvipsnames]{xcolor}
\usepackage{makecell, cellspace, caption}
\usepackage{enumitem}
\usepackage{amsmath,amssymb,graphicx,dblfloatfix,mathtools}
\usepackage{relsize}
\usepackage{booktabs}
\newcommand{\STAB}[1]{\begin{tabular}{@{}c@{}}#1\end{tabular}}
\usepackage{empheq}
\usepackage{xcolor}
\newcommand{\boxedeq}[2]{\begin{empheq}[box={\fboxsep=6pt\fbox}]{align}\label{#1}#2\end{empheq}}
\usepackage[T1]{fontenc}
\newcommand{\bigF}{\scalebox{1.2}{$\mathcal{F}$}}
\usepackage[backend=biber, sorting=none, style=numeric-comp]{biblatex}
\usepackage{xcolor}

\newcommand*{\colorboxed}{}
\def\colorboxed#1#{%
  \colorboxedAux{#1}%
}
\newcommand*{\colorboxedAux}[3]{%
  \begingroup
    \colorlet{cb@saved}{.}%
    \color#1{#2}%
    \boxed{%
      \color{cb@saved}%
      #3%
    }%
  \endgroup
}

\usepackage{empheq}

\usepackage{capt-of}
\usepackage[table,x11names]{xcolor}
\definecolor{flashwhite_v2}{rgb}{0.92, 0.92, 0.98}
\definecolor{azure}{rgb}{0.94, 1.0, 1.0}
\newcommand{\CC}[1]{\cellcolor{flashwhite_v2}}
\usepackage{makecell}
\usepackage[hyperfootnotes=false]{hyperref}

\title{Deep Neural Mel-Subband Beamformer for In-car Speech separation}
\name{Vinay Kothapally, Yong Xu, Meng Yu, Shi-Xiong Zhang, Dong Yu}
\address{Tencent AI Lab, Bellevue, WA, USA}

\begin{document}
\ninept
\setlength{\abovedisplayskip}{1pt}
\setlength{\belowdisplayskip}{5pt}

\maketitle

\begin{abstract}
While current deep learning (DL)-based beamforming techniques have been proved effective in speech separation, they are often designed to process narrow-band (NB) frequencies independently which results in higher computational costs and inference times, making them unsuitable for real-world use. In this paper, we propose DL-based mel-subband spatio-temporal beamformer to perform speech separation in a car environment with reduced computation cost and inference time. As opposed to conventional subband (SB) approaches, our framework uses a mel-scale based subband selection strategy which ensures a fine-grained processing for lower frequencies where most speech formant structure is present, and coarse-grained processing for higher frequencies. In a recursive way, robust frame-level beamforming weights are determined for each speaker location/zone in a car from the estimated subband speech and noise covariance matrices. Furthermore, proposed framework also estimates and suppresses any echoes from the loudspeaker(s) by using the echo reference signals. We compare the performance of our proposed framework to several NB, SB, and full-band (FB) processing techniques in terms of speech quality and recognition metrics. Based on experimental evaluations on simulated and real-world recordings, we find that our proposed framework achieves better separation performance over all SB and FB approaches and achieves performance closer to NB processing techniques while requiring lower computing cost.
\end{abstract}



\vspace{-1em}
\section{Introduction}
\label{sec:intro}
\vspace{-0.5em}
A combination of deep learning (DL) and voice-activated technologies are rapidly advancing in the automotive industry, enabling improved human-machine interaction (HMI) and advanced intelligent assistance \cite{car_01}. Such systems are challenging to develop for several reasons: (i) they must demonstrate multidisciplinary expertise in technologies such as speech enhancement (SE)\cite{car_se,car_se_asr,SkipConvNet,SkipConvGAN,ComplexAttention,SE_and_SAD}, automatic speech recognition (ASR) \cite{car_asr,caroselli2022cleanformer}, speaker identification (SID) \cite{car_sid}, speech separation \cite{o2022universally,o2021conformer}, etc., (ii) their performance must be robust against a wide range of noise conditions encountered during the drive, and (iii) they must operate with the limited in-car computational power and possess a faster inference time for quick response times. The scope of this study is limited to enhancing in-car speech separation abilities with reduced computation costs and faster inference.


Currently, most state-of-the-art deep learning based speech separation systems are built using mask-based minimum variance distortion-less response (MVDR) beamformer \cite{mvdr,adlmvdr,mcmf, grnnbf,jaecbf} and its variants which are capable of spatially filtering speech with lesser non-linear distortions compared to any purely "black-box" neural network (NN) based approaches \cite{mc_td_enh,mc_gnn_enh,mc_nn_enh,mc_attn_enh,haubner2022deep}. However, these conventional mask based beamformer approaches always suffer from high residual noise.  Our recent study on all-deep-learning MVDR (ADL-MVDR) \cite{adlmvdr} and generalized recurrent neural network beamformer (GRNN-BF) \cite{grnnbf} solve the prior issue by computing frame-level beamforming weights that are suitable for streaming scenarios, such as in-car applications. Likewise, we recently proposed deep learning-based Joint AEC and beamforming techinque (JAEC-BF) \cite{jaecbf} an advancement to aforementioned beamformers to handle non-linear echoes from the loudspeakers received as feedback with the use of joint statistics of microphone and echo reference signals.

Despite their effectiveness, all-deep-learning beamforming techniques have a slower inference time and require high computational power, since they are narrow-band (NB) systems \cite{nb_mc_se, mc_nb_separation,mc_conf_separation}, which process each frequency within speech independently. In contrast, full-band (FB) systems that process all frequencies simultaneously suffer with reduced overall performance. Xiaofei et al. \cite{nb_mc_se}, proposed that the narrow band beamforming could avoid the over-fitting problem due to the frequency dependencies and generalize well to the unseen real-world testing. To this end, we proposed an end-to-end subband spatio-temporal RNN beamformer for in-car speech separation with the aim of achieving faster inference with reduced computation cost, see Fig.\ref{Network}. Furthermore, we improve the performance of the proposed subband (SB) beamformer by introducing a novel mel-scale based subband selection strategy. The proposed subband selection approach provides the beamformer with a higher-resolution for lower-frequencies rich in speech content and a lower-resolution for higher-frequencies adequate to capture spatial cues. We conduct extensive tests to present performance improvements achieved by proposed speech separation framework on a wide variety of simulated and actual in-car recordings in comparison with other multi-channel full-band and narrow-band approaches.


\vspace{-1em}
\section{Problem formulation}
\label{sec:problem}
\begin{figure*}[htb!]
  \centering
  \centerline{\includegraphics[width=\textwidth]{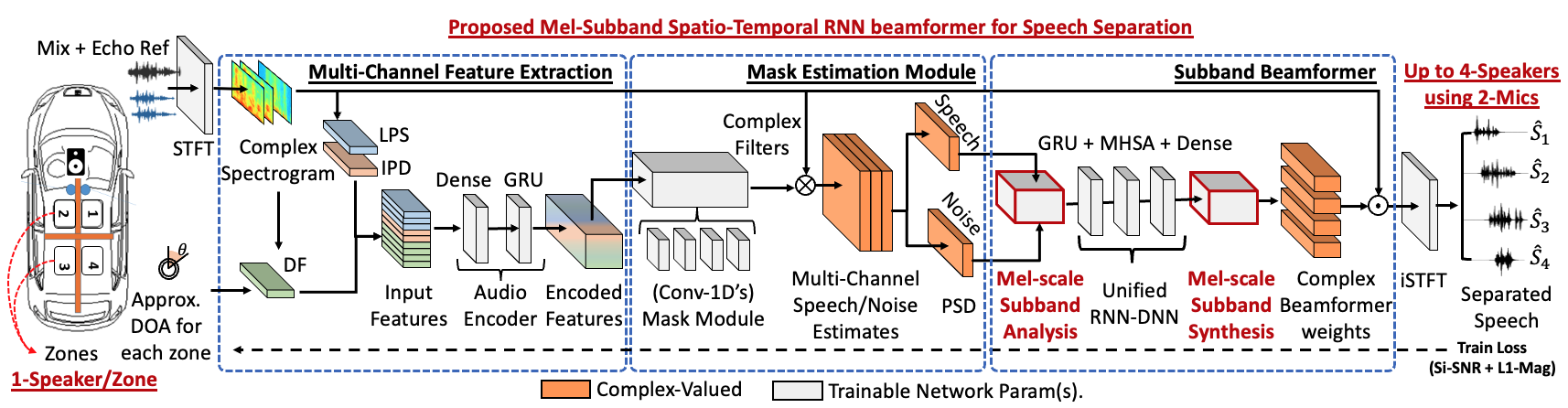}}
  \vspace{-1em}
  \caption{An overview of the proposed all deep learning-based speech separation system for in-car applications}
  \label{Network}
  \vspace{-1em}
\end{figure*}

\vspace{-0.5em}
We consider the problem of in-car `$N$'-speaker speech separation using `$M$'-channel microphone array. Let $s_{i}(l)$ and $x(l)$ represent the clean speech from $i^{th}$-speaker and in-car loudspeaker signal respectively. Assuming only one speaker is permitted to move within a designated zone while speaking, see Fig.\ref{Network} for speaker zones. Speech captured by the microphone array, $\mathbf{y}(l)$ (termed as ``mix") at time sample `$l$' can be represented as,
\vspace{-0.2em}
\begin{equation} 
\begin{aligned}[b]
\mathbf{y}(l) & = \smashoperator{\sum_{i=0}^{N{-}1}}\mathbf{h}_{i}(l)\ast s_{i}(l) + \mathbf{h}_{x}(l) \ast f_{\text{NL}}(x(l)) + \mathbf{v}(l)
\label{eq:mixture}
\end{aligned}
\end{equation}

\noindent where, `$l$' is the sample index, `$f_{\text{NL}}$' represents non-linearity introduced by in-car loudspeaker(s), $\mathbf{h}_{i}(l)$ and $\mathbf{h}_{x}(l)$ are $M$-channel room impulse responses (RIRs) from $i^{th}$-speaker and the loudspeaker locations to the microphone array, `$\ast$' denotes convolution, and $\mathbf{v}(l)$ represents background noise. This study aims at developing a DL-based end-to-end speech separation system $(\mathlarger{\boldsymbol \Psi_\mathrm{proposed}})$ to effectively separate out all speakers' speech simultaneously suppressing background noise and echoes from loudspeaker with lower computational costs and faster inference on CPU, see Eq.\eqref{eq:model}. Here `$(\cdot)^T$' represents vector transpose and $\hat{\mathbf{s}}^{r}$ vectorizes all $\hat{s}^{r}_i,\:\forall i{\in}[0, N{-}1]$. Furthermore, we emphasize that the proposed separation system is limited to estimating reverberant clean speech, $\hat{s}^{r}_{i}(l) = \mathbf{h}_{i}(l){\ast}s_{i}(l)$ but not the anechoic speech, $\hat{s}_{i}(l)$.

\vspace{-0.5em}
\boxedeq{eq:model}{\hat{\mathbf{s}}^{r}(l) = \boldsymbol \Psi_\mathrm{proposed}([\mathbf{y}(l),\; x(l)]^T)}

\vspace{-1em}
\section{Background and Motivation}
\vspace{-0.5em}

The following section discusses DL-based narrow-band (NB), full-band (FB), and subband (SB) techniques to help outline our mel-scale subband framework. Let us consider mixture, clean in-car loudspeaker and reverberant clean speech signals of $i^{th}$-speaker in Eq.\eqref{eq:mixture} expressed in short-time fourier transform (STFT) domain as, $\mathbf{Y}_{t,f}$, $X_{t,f}$ and $\mathbf{S}^{r}_{t,f}$ where $t{\in}[0,T{-}1]$ and $f{\in}[0,F{-}1]$ represent frame-index and frequency bin respectively. \\

\noindent{\textbf{Narrow-band (NB) processing: }}An NB-system $(\mathlarger{\boldsymbol \Psi_\mathbf{NB}})$ estimates reverberant clean speech of all speakers $\hat{\mathbf{S}}^{r}_{t,f}$ by processing each frequency bin individually, as illustrated below:

\vspace{-0.5em}
\begin{equation}
  \colorboxed{black}{
    \hat{\mathbf{S}}^{r}_{f}(t) = \boldsymbol \Psi_\mathbf{NB}([\mathbf{Y}_f(0{:}t),\; X_f(0{:}t)]^T) \;\;  \forall\:f\:{\in}\:[0, F{-}1]
  }
\label{eq:NB}
\end{equation}


\noindent where $\hat{\mathbf{S}}^{r}_{f}(t),\mathbf{Y}_f(t), X_f(t)$ represent a time sequence for one frequency bin. The network $(\mathlarger{\boldsymbol \Psi_\mathbf{NB}})$ is shared across all frequencies since NB-systems learn a unique function applicable across all frequency bins. Theses systems provide the best spatial suppression \cite{NB_enh} by capturing challenging speaker patterns across frequencies. However, the network must process each frequency bin independently, which increases computations and inference time.\\


\noindent{\textbf{Full-band (FB) processing: }}Unlike an NB-system, a FB-system $(\mathlarger{\boldsymbol \Psi_\mathbf{FB}})$ process full-band spectrum at once to estimate reverberant clean speech $\hat{\mathbf{S}}^{r}_{t,f}$ for all speakers, see Eq.\eqref{eq:FB}. Here $\mathbf{Y}(t),X(t)$, and $\mathbf{\hat{S}}^{i}(t)$ represents a vector with all frequencies flattened and concatenated for $\mathbf{Y}_{t,f}$, $X_{t,f}$ and $\mathbf{S}^{r}_{t,f}$ signals respectively.


\vspace{-0.5em}
\begin{equation}
  \colorboxed{black}{
    \mathbf{\hat{S}}^{r}(t) = \boldsymbol \Psi_\mathbf{FB}([\mathbf{Y}(0{:}t),\; X(0{:}t)]^T) 
  }
\label{eq:FB}
\end{equation}

\noindent As FB-based systems estimate full-band spectrum in a single processing step, they require fewer computations. However, they have the overfitting issue due to the concatenated frequencies and generalize worse to real world scenarios \cite{NB_enh}.\\



\noindent{\textbf{Subband (SB) processing: }} An SB-system is widely used to address the trade-off between performance and computational cost by helping to minimize the overall model size. This is achieved by splitting a full-band spectrum into several contiguous frequency bands, which are independently processed like an NB-system. Recent study \cite{dccrn++} has shown improved performance using SB split and merge neural layer modules. However, the SB technique in these frameworks splits frequencies uniformly, creating sub-bands with equal bandwidths. Considering that spectral content from several frequency bins are combined within a band, this may cause problems in bands corresponding to low frequencies where speech is highly active, compromising system's overall performance. This motivates us to develop mel-scale SB-system that splits a full-band spectrum into non-uniform bands where fewer frequency bins are combined within a band at low frequencies compared to high-frequency bands.

\vspace{-1em}
\section{Proposed Mel-Subband beamformer}
\label{sec:typestyle}
\vspace{-0.5em}
The overall architecture of the proposed mel-scale subband spatio-temporal RNN beamformer (SRNN-BF) is depicted in Fig.\ref{Network} which comprises of three modules: (i) a multi-channel feature extraction, (ii) a mask estimation module, and (iii) a subband beamformer jointly trained in an end-to-end approach. As stated in Eq.\eqref{eq:model}, the proposed model is provided with stacked `$M$'-channel mixture and single-channel in-car loudspeaker reference signals $\mathbf{y}(l)$. The audio samples are transformed to frequency domain, $\mathbf{Y}_{t,f}$ and $X_{t,f}$ using 1-D convolution layers that employ STFT operations. To this end, we extract log power spectrum (LPS), inter-channel phase difference (IPD), and directional angle features (DF) \cite{grnnbf} and merge them using convolutional and recurrent layers to server as input features. \\

\label{ssec:masking} 
\noindent{\textbf{4.1. Mask estimation module: }}The mask estimation module uses the encoded features to predict speech and noise masks, which indicates the time-frequency (TF) bins where the target source or noise is dominant. These estimated masks are then used to estimate the spatial covariance matrices (SCMs) of the target speech and noise, which are required to compute the beamformer coefficients. However, our previous studies on neural beamformers have shown that complex-valued ratio filters (cRFs) \cite{deep_filters} provide better estimates of SCMs for multi-channel speech and noise. Additionally, our recent work in \cite{jaecbf} showed that joint SCMs estimated using both mixture and echo reference signals can produce better beamforming solution, especially in presence of acoustic echo. Therefore, we use one-dimensional (1-D) convolution layers in our proposed network to estimate `$\mathrm{cRF_S}$' and `$\mathrm{cRF_Z}$' for mixture and in-car loudspeaker echo reference signals to estimate joint SCMs for speech and noise, see Fig.\ref{Network}. As demonstrated in Eq.\eqref{eq:cRF}, we employ $\mathrm{cRF_{S}}$ on mixture and in-car loudspeaker echo signals to compute multi-channel speech estimate, $\mathbf{\widetilde{S}}_{t,f}{\in}\mathbb{C}^{((M{+}1)\times T\times F)}$. 

\vspace{-0.5em}
\begin{equation}
  \colorboxed{black}{
    \mathbf{\widetilde{S}}_{t,f}=\smashoperator{\sum_{\tau \in [-K,K]}}\mathrm{cRF_{S}}(t,f,\tau)\ast[\mathbf{Y}_{t{+}\tau,f}, X_{t{+}\tau,f}]^T
  }
\label{eq:cRF}
\end{equation}

Here `$\tau$' represents the filter-taps of cRF filters. Next, the multi-channel speech estimate $\mathbf{\widetilde{S}}_{t,f}$ are used to compute frame-wise speech SCMs $\mathbf{\Phi_{\widetilde{S}\widetilde{S}}}(t,f){\in}\mathbb{C}^{(T\times F\times (M{+}1)\times (M{+}1))}$, see Eq.(\ref{eq:SCMs}). 

\vspace{-0.5em}
\begin{equation}
  \colorboxed{black}{
    \mathbf{\Phi_{\widetilde{S}\widetilde{S}}}(t,f)= \mathrm{LayerNorm}\big(\mathbf{\widetilde{S}}_{t,f}\mathbf{\widetilde{S}}^H_{t,f}\big)
  }
\label{eq:SCMs}
\end{equation}

Similar computations are carried out using $\mathrm{cRF_{Z}}$ to estimate multi-channel noise estimates, $\mathbf{\widetilde{Z}}_{t,f}$ and compute frame-wise noise spatial covariance matrix $\mathbf{\Phi_{\widetilde{Z}\widetilde{Z}}}(t,f)$. As mentioned in \cite{grnnbf}, we use layer normalization on both speech and noise SCMs to achieve better performance and faster convergence. Furthermore, since the clean in-car loudspeaker signals are used to jointly model the cross-correlations among the mixture and in-car loudspeaker signals, this aids the network with echo suppression \cite{jaecbf}.\\

\noindent{\textbf{4.2. Sub-band beamformer: }} Similar to \cite{grnnbf}, the proposed SB-beamformer takes the estimated speech \& noise SCMs as inputs to predict the complex-valued beamforming weights. To this end, we flatten and concatenate real and imaginary parts of these SCMs to form a `$D_\mathrm{in}$'-dimensional vectors $\mathbf{\Phi_{\widetilde{S}\widetilde{S}}}$ and $\mathbf{\Phi_{\widetilde{Z}\widetilde{Z}}}{\in}\mathbb{R}^{(T\times F\times D_\mathrm{in})}$ for each TF bin. Using the below described mel-scale subband analysis module, we split full-band SCMs into `$K$' non-uniform sub-bands.\\

\noindent \textbf{Mel-scale subband analysis: }
We begin by splitting the full-band spectrum into $K$-subbands on traditional mel-scale, such that each subband consists of $f^{sub}_k$ number of frequencies. Then, we employ a total of `$K$' 2-D convolutional filters as learnable analysis filters independently of each subband-`$k$' $\forall \:k\:{\in}\:[0,K{-}1]$, to project non-uniform subbands with varying number of frequency bins to `$E$'-dimensional space for further processing. Thus, a `$k^{th}$' subband of speech SCM ($\mathbf{\Phi}^{k}_\mathbf{\widetilde{S}\widetilde{S}}$) can be expressed as:



\begin{align}
\boxed{
\resizebox{0.88\linewidth}{!}{$
    \begin{array}{ccc}
        \mathbf{\Phi}^{k}_\mathbf{\widetilde{S}\widetilde{S}}(t,f) = \big[\mathbf{\Phi_{\widetilde{S}\widetilde{S}}}(t,f_{k}),.., \mathbf{\Phi_{\widetilde{S}\widetilde{S}}}(t,f_{k{+}1}) \big]^T \:{\in}\:\mathbb{R}^{(f^{sub}_k\times T\times D_\mathrm{in})} \\ \\
        \mathbf{\Phi}^{k}_\mathbf{\widetilde{S}\widetilde{S}}(t,e) = \mathrm{\mathrm{Conv2D}}_k\big(\mathbf{\Phi}^{k}_\mathbf{\widetilde{S}\widetilde{S}}(t,f)\big) \quad {\in}\;\mathbb{R}^{(E\times T\times  D_\mathrm{in})} 
    \end{array}
$}
}
\label{eq:subband_scm1}
\end{align}



Similar computations are carried out for noise SCM to produce subband noise SCM, $\mathbf{\Phi}^k_\mathbf{\widetilde{Z}\widetilde{Z}}(t,e)$. We now feed the SB speech and noise SCMs to a unified RNN-DNN ($\bigF_\textrm{RD}$) layers which are shared across all sub-bands to predict SB-level multi-frame beamforming weights for each speaker as, 

\vspace{-0.5em}
\begin{equation}
  \colorboxed{black}{
    \mathrm{\textbf{w}}^k_{t,e,\tau} = \bigF_\textrm{RD}\big( \big[ \mathbf{\Phi}^{k}_\mathbf{\widetilde{S}\widetilde{S}}(0{:}t,e), \mathbf{\Phi}^{k}_\mathbf{\widetilde{Z}\widetilde{Z}}(0{:}t,e)\big]\big)
  }
\label{eq:rnn_dnn}
\end{equation}

\noindent where `$\tau$' represents the filter taps of the estimated beamformer weights, and  $\mathrm{\textbf{w}}^k_{t,e}$ represents a vector containing the subband multi-frame beamforming weights for each zone in the car. Furthermore, a multi-head self-attention (MHSA) \cite{vaswani2017attention}, used to dynamically adapt these weights of each speaker. Similar to \cite{mcmf}, the unified RNN-DNN in our proposed subband approach is a multi-channel and multi-frame (MCMF) spatio-temporal recurrent beamformer.\\ 



\begin{table*}[htb!]
    \begin{minipage}[b]{0.63\linewidth}
        \centering
        \renewcommand{\arraystretch}{2.0}
        \resizebox{\linewidth}{!}{
            \begin{tabular}{l|l|l|c|c|cccc|c}\hline
                \multicolumn{3}{l|}{\multirow{2}{*}{\textbf{\Large{DL-based Systems}}}} & \multirow{2}{*}{\textbf{\Large{\#Param}}} & \multirow{2}{*}{\makecell{ \textbf{\Large{\#GMACs}}\\\textbf{\large{(per sec.)}}}} &  \multicolumn{4}{c|}{\textbf{\Large{Simulated Data}}} &   \multicolumn{1}{c}{\textbf{\Large{Real Data}}} \\ 
                \multicolumn{3}{l|}{} & \multicolumn{1}{l|}{} & \multicolumn{1}{l|}{} & \textbf{\Large{PESQ}} 
                & \textbf{\Large{SiNSR}} & \textbf{\Large{SDR}} & \textbf{\Large{WER (\%)}} & \textbf{\Large{WER (\%)}} \\\hline
                
                \multicolumn{3}{l|}{\Large{No processing}} & \Large{-} & \Large{-} & \Large{1.566} & \Large{-4.523} & \Large{-4.374} & \Large{ > 100} & \Large{> 100}\\ \hline\hline
                
                \multirow{3}{*}{\STAB{\rotatebox[origin=c]{90}{\Large{FB}}}} & \multicolumn{2}{l|}{\Large{LSTM + MHSA}\cite{lstm_SS}}   & \Large{22.98} & \Large{1.37} & \Large{2.362} & \Large{6.924} & \Large{7.948} & \Large{23.10} & \Large{30.33} \\ 
                
                & \multicolumn{2}{l|}{\Large{ConvTasNet}}                                                               & \Large{23.65} & \Large{1.12} & \Large{2.362} & \Large{4.109} & \Large{4.257} & \Large{23.99} & \Large{31.29}\\ 
                
                & \multicolumn{2}{l|}{\CC{20}\Large{\textbf{GRNNBF}} \cite{grnnbf}}                       & \CC{20}\Large{\textbf{23.75}} & \CC{20}\Large{\textbf{1.32}} & \CC{20}\Large{\textbf{2.438}} & \CC{20}\Large{\textbf{7.582}} & \CC{20}\Large{\textbf{8.542}} & \CC{20}\Large{\textbf{21.37}} & \CC{20}\Large{\textbf{31.27}} \\\hline
    
                \multirow{8}{*}{\STAB{\rotatebox[origin=c]{90}{\Large{Subband Processing}}}} & 
                \multirow{4}{*}{\STAB{\rotatebox[origin=c]{90}{\Large{Traditional}}}} & \Large{\#SB 08}                  & \Large{5.81} & \Large{1.47} & \Large{2.488} & \Large{7.426} & \Large{8.406}   & \Large{22.54} & \Large{26.03} \\
               
                &  & \Large{\#SB 16}                                                                                     & \Large{5.07} & \Large{2.16} & \Large{2.630} & \Large{8.160} & \Large{9.101}   & \Large{12.38} & \Large{15.62} \\
                
                &  & \Large{\#SB 32}                                                                                     & \Large{4.70} & \Large{3.56} & \Large{2.825} & \Large{9.245} & \Large{10.181}  & \Large{8.04}  & \Large{11.23} \\

                &  & \CC{20}\Large{\textbf{\#SB 64}}                                                       & \CC{20}\Large{\textbf{6.35}} & \CC{20}\Large{\textbf{4.52}} & \CC{20}\Large{\textbf{2.784}} & \CC{20}\Large{\textbf{9.257}} & \CC{20}\Large{\textbf{10.242}}  & \CC{20}\Large{\textbf{7.23}}  & \CC{20}\Large{\textbf{11.03}} \\ \cline{2-10}

                & \multirow{4}{*}{\STAB{\rotatebox[origin=c]{90}{\Large{Proposed}}}}  & \Large{\#SB 08}                  & \Large{6.22} & \Large{3.68} & \Large{2.483} & \Large{7.612} & \Large{8.238}   & \Large{19.46} & \Large{19.04} \\
                
                &  & \Large{\#SB 16}                                                                                     & \Large{5.28} & \Large{3.29} & \Large{2.759} & \Large{8.763} & \Large{9.805}   & \Large{10.01} & \Large{12.22} \\
                
                &  & \Large{\#SB 32}                                                                                     & \Large{4.81} & \Large{4.15} & \Large{2.845} & \Large{9.382} & \Large{10.381}  & \Large{7.85}  & \Large{9.83} \\

                &  & \CC{20}\Large{\textbf{\#SB 64}}                                                       & \CC{20}\Large{\textbf{6.67}} & \CC{20}\Large{\textbf{4.57}} & \CC{20}\Large{\textbf{3.024}} & \CC{20}\Large{\textbf{10.712}} & \CC{20}\Large{\textbf{11.632}} & \CC{20}\Large{\textbf{5.53}}  & \CC{20}\Large{\textbf{9.35}} \\ \hline

                \multirow{3}{*}{\STAB{\rotatebox[origin=c]{90}{\Large{NB}}}} & \multicolumn{2}{l|}{\Large{MVDR}\cite{adlmvdr}}  & \Large{4.55} & \Large{0.05} & \Large{2.161} & \Large{1.352} & \Large{3.849} & \Large{77.37} & \Large{83.98} \\
                & \multicolumn{2}{l|}{\Large{TV-MVDR \cite{higuchi2018frame}}}                                                                   & \Large{4.56} & \Large{0.67} & \Large{2.272} & \Large{4.127} & \Large{5.183} & \Large{32.66} & \Large{40.78} \\

                & \multicolumn{2}{l|}{\CC{20}\Large{\textbf{GRNNBF}} \cite{grnnbf}}   & \CC{20}\Large{\textbf{4.33}} & \CC{20}\Large{\textbf{22.46}} & \CC{20}\Large{\textbf{3.057}} & \CC{20}\Large{\textbf{11.309}} & \CC{20}\Large{\textbf{12.245}} & \CC{20}\Large{\textbf{5.21}} & \CC{20}\Large{\textbf{8.69}} \\\hline\hline

                \multicolumn{3}{l|}{\Large{Reference Rev. Clean}}                                                       & \Large{-} & - & \Large{4.50} & \Large{$\infty$} & \Large{$\infty$} & \Large{1.32} & \Large{-NA-}\\ \hline
            \end{tabular}}
        \caption{Experimental results for different DL-based frameworks across objective \\evaluation metrics, computation cost (GMACs), and word error rate (WER).}
        \label{table:metrics}
    \end{minipage}\hfill
    \begin{minipage}[b]{0.34\textwidth}
        \centering
        \includegraphics[height=65mm, width=56mm]{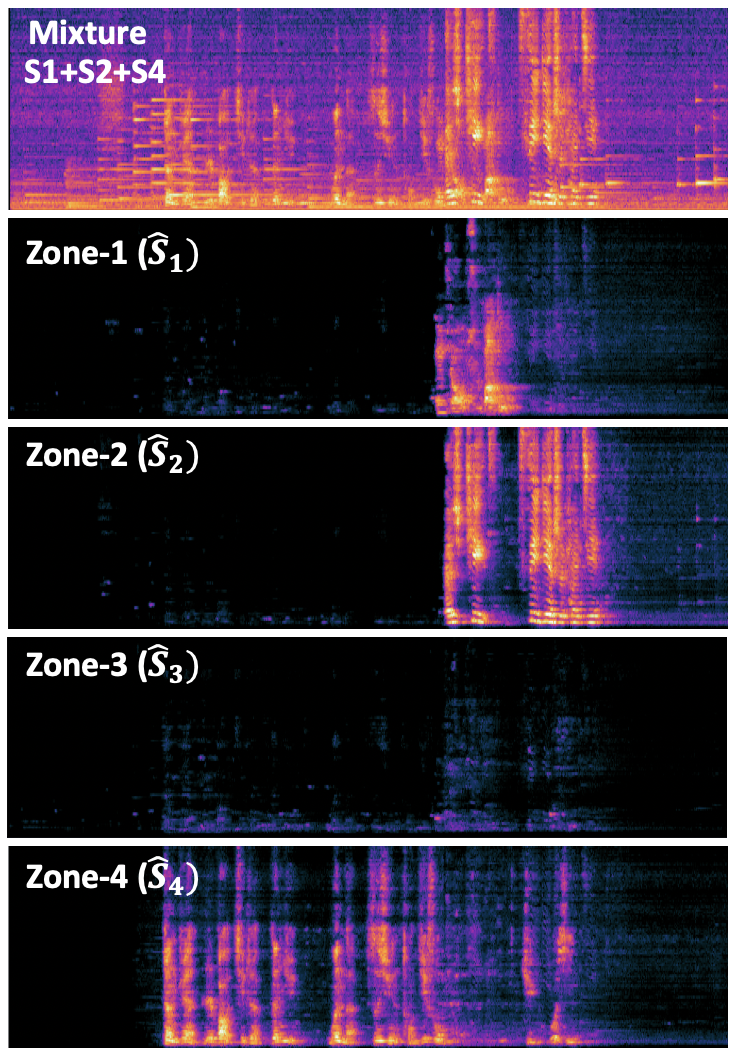} \vspace{0.5em} 
        \captionof{figure}{3-Speaker (speaker-3 absent) noisy mixture and separated speech spectrograms estimated by proposed 64-subband beamformer.}
        \label{fig:spec}
    \end{minipage}
    \vspace{-2.0em}
\end{table*}


\noindent \textbf{Mel-scale subband synthesis:} The estimated SB-level beamformer weights are converted to full-band by employing 2-D convolutional learnable subband synthesis filters independently for each subband-`$k$', $\forall \:k\:{\in}\:[0,K{-}1]$. Later, the outputs are concatenated along frequency axis to retrieve full-band MCMF beamformer weights. 

\begin{align}
\boxed{
\resizebox{0.88\linewidth}{!}{$
    \begin{array}{cc}
        \mathrm{\textbf{w}}^k_{t,f,\tau} = \mathrm{Conv2D}_k\big(\mathrm{\textbf{w}}^k_{t,e,\tau}\big) \quad {\in}\;\mathbb{C}^{(f^{sub}_k\times T\times D_\mathrm{out})}  \\
        \mathrm{\mathbf{W}_\mathrm{proposed}}(t,f,\tau) = Concat\big[\mathrm{\textbf{w}}^0_{t,f,\tau}, \mathrm{\textbf{w}}^1_{t,f,\tau} ..,  \mathrm{\textbf{w}}^{K{-}1}_{t,f,\tau}\big]^T 
    \end{array}
$}
}
\label{eq:synthesis}
\end{align}



We compute the speech spectrum for all speakers $\hat{\mathbf{S}}^{r}_{t,f}$ by filtering mixture and in-car loudspeaker signals using complex-valued MCMF-weights estimated by the proposed SB beamformer as, 

\vspace{-0.5em}
\begin{equation}
  \colorboxed{black}{
    \hat{\mathbf{S}}^{r}_{t,f}=\smashoperator{\sum_{\tau \in [-K,K]}}\:\mathrm{\mathbf{W}_\mathrm{proposed}}(t,f,\tau)^H[\mathbf{Y}_{t{+}\tau,f}, X_{t{+}\tau,f}]^T
  }
\label{eq:beamform}
\end{equation}


Finally, $\hat{\mathbf{S}}^{r}_{t,f}$ is transformed to time domain, $\hat{\mathbf{s}}^{r}(l)$ using 1-D convolution layers that employ the inverse-STFT operation.

\vspace{-1em}
\section{Dataset and Experimental Setup}
\label{sec:dataset}
\vspace{-0.5em}
\subsection{Dataset}
\label{ssec:dataset}
\vspace{-0.5em}
We simulate multi-speaker in-car dataset using AISHELL-2 \cite{aishell_2} and AEC-Challenge \cite{AEC_sim1} corpus. We generate a total of 10k multi-channel RIRs with random vehicle dimensions using image-source method. Each multi-channel RIR is a set consisting of RIRs from four speakers (one in each zone), loudspeaker, and background noise locations to 2-channel linear microphone array measuring 11.8 cm in length. The reverberation time (RT$_{60}$) ranges between [0,0.6s] across configurations. We randomly select RIRs to simulate in-car dataset. In addition, we use clean and nonlinear distortion functions from AEC-Challenge \cite{AEC_sim1} to simulate the loudspeaker signals. The nonlinear distortions include, but are not limited to: (i) clipping the maximum amplitude, (ii) using a sigmoidal function \cite{AEC_sim2}, and (iii) applying learned distortion functions. In addition, we include diffused noise with SNRs ranging from [-40,15] dB and  signal to echo ratio (SER) from [-10,10] dB. A total of 180K, 7.5K, and 2K utterances are generated for the `\textit{Train}', `\textit{Dev}', and `\textit{Test}' datasets.

\vspace{-0.5em}
\subsection{Experimental Setup}
\label{ssec:expsetup}
\vspace{-0.5em}
A 512-point STFT is employed with 32 ms Hann window and 16 ms step size to extract complex spectra for mixture and far-end signals. All systems in the study are trained on 4-second chunks with the Adam optimizer and a batch size varied within 12 and 24 to maximize the time-domain scale-invariant source-to-noise ratio (Si-SNR) \cite{SiSNR} and minimize the frequency-domain mean square error (MSE) between estimated speaker speech and corresponding reverberant clean speech signals, both of which are equally weighted. Initial learning rate is set to 1e-4 with a gradient norm clipped with max norm 10. All subband systems in this study are designed to have less than 5M parameters to be memory efficient and trained for 30 epochs. We use of kernel-size of $(1{\times}1)$ and $(3{\times}1)$ for all 1-D \& 2-D convolutions within the network. The estimated cRFs in the proposed system are filters with 3 taps. Likewise, the estimated multi-channel and multi-frame beamformer are filters with 5 taps. In this study, we compare our proposed method to several NB, FB and SB-based baseline systems which include: (i) NB: traditional MVDR \cite{mvdr}, time-variant MVDR, GRNNBF \cite{grnnbf}, a robust NN-based beamformer, (ii) FB: widely used LSTM and MHSA module \cite{lstm_SS}, GRNNBF \cite{grnnbf}, frequency domain ConvTasNet, and (iii) SB: GRNNBF with conventional subband approach.


\vspace{-1em}
\section{Results and Discussion}
\label{sec:prior}
\vspace{-1em}

We compare the performance of the proposed neural mel-subband beamformer system to conventional subband techniques and other FB and NB-based beamforming frameworks for in-car speech separation task using perceptual quality measure such as PESQ, as well as objective metrics such as Si-SNR, signal-to-distortion (SDR) ratio on simulated `$Test$' set in Table \ref{table:metrics}. Furthermore, we use a general-purpose mandarim speech recognition Tencent API \cite{TencentASR} to test the ASR performance of all frameworks on simulated and real-worl recordings by computing the word error rate (WER). We also compare the computational complexity and cost required for each framework by computing multiply-accumulate additions (MACs) for one second of audio samples. For ease of understanding, Table.\ref{table:metrics} is divided into three sections, each listing the objective scores and WER achieved for FB, SB, and NB frameworks. The frameworks that achieve lowest WER in each category are highlighted.\\

\noindent{\textbf{Proposed Mel subband vs Traditional subband processing: }} We compare the performance of our proposed mel subband beamformer with the traditional subband approach for 8, 16, 32, and 64 subband settings. From Table.\ref{table:metrics}, we see that our proposed method consistently outperforms the traditional subband across all objective measures as well as in back-end speech recognition task. For example, our method improves PESQ, Si-SNR, and SDR by 8.62\%, 16.35\%, and 13.57\% over the traditional approach using 64 sub-bands. Furthermore, we see that our approach achieves a relative improvement of 30.74\% and 17.96\% in WER over the traditional approach when tested on simulated and real in-car speech recordings with minimal to no increase in computing cost, i.e., GMACs: 4.57 vs 4.52 (1.1\% relative increase). Likewise, we see that our approach achieves a relative improvement of \{20.25, 20.45, 7.41, 19.37\}\% in averaged WER over the traditional approach using 8, 16, 32, and 64 sub-bands respectively. These findings suggest that the proposed mel subband processing strategy improves the traditional subband approach by preserving lower frequencies with most speech patterns.\\

\noindent{\textbf{Proposed Mel subband vs [NB \& FB] processing: }}It is known to all that NB-systems are proven to be the most efficient in several speech tasks, such as speech enhancement and speech separation. As we demonstrate in our experiments, the NB-variant of our previously proposed GRNNBF system outperforms all systems. For instance, NB-based GRNNBF achieves a 73.91\% relative improvement in WER over simulated and real in-car recordings compared to its FB-counterpart. Likewise, NB-based GRNNBF achieves a relative improvement of 24.57\% over traditional 64 subband system. While traditional 64 subband systems $5{\times}$ faster and computationally inexpensive in comparison to NB-based GRNNBF, it is not capable of preserving the performance in terms of both perceptual quality and speech recognition task. In contrast to the traditional subband approach, our approach is able to better preserve the performance achieved by NB-system in overall speech quality and speech recognition while reducing computing costs by $5{\times}$ times and the overall CPU memory usage. For example, our 64 subband approach achieves PESQ: 3.024 vs 2.057, WER (simulated data): 5.53 vs 5.21, and WER (real data): 9.35 vs 8.69. Likewise, our proposed method significantly outperforms all FB-based systems over all objective metrics and WER, see Table. \ref{table:metrics} with an average $3{\times}$ increased computing cost. For example, relative improvement of \{24.03, 41.28, \& 72.11\}\% in PESQ, Si-SNR, and average WER over FB-based GRNNBF system. Based on these findings, we strongly suggest that the proposed mel subband beamformer can preserve NB-system performance while reducing computation costs. \\

\noindent{\textbf{In-Car Speaker Separation Demo: }}We present a in-car scenario with three speakers present in zones-1,2 \& 4 respectively. Fig.\ref{fig:spec} shows the spectrograms for mixture and separated speech signals estimated using our proposed 64-band mel subband beamformer. It is evident from the separated speech spectrograms that the proposed mel subband beamformer can efficiently separate speech from up to four speakers (one in each zone) in a car environment using only two microphones, which is an extremely challenging task. Readers can find several audio samples from simulated and real test set processed by various system mentioned in the study at \href{https://vkothapally.github.io/Subband-Beamformer}{\color{blue}\textup{https://vkothapally.github.io/Subband-Beamformer}}.  

\vspace{-1.3em}
\section{Conclusion}
\label{sec:conclusion}
\vspace{-1em}
To conclude, we propose a subband beamformer for in-car speech separation preserves the ASR performance while reducing CPU processing time. First, the proposed framework divides the full spectrum into non-uniform sub-bands based on mel-scale to ensure spectral information embedded in lower frequencies, where speech patterns are most prevalent are preserved. Furthermore, the learnable neural analysis filters within the proposed framework enhance the subband representations. By combining non-uniform band splitting strategy with learnable filters, we are able to preserve speech separation and recognition performance at a lower cost. The proposed mel-subband is compared to conventional subband approaches and other widely used neural beamforming frameworks for speech separation. We demonstrated through several subband configurations that our proposed method achieves better WER and objective scores. 


\vfill\pagebreak

\section{References}
\printbibliography[heading=none]

\end{document}